\documentclass[aps,prl,twocolumn,showkeys,floatfix,nofootinbib,superscriptaddress]{revtex4-2}

\usepackage{float}
\usepackage{xcolor}
\usepackage{graphicx}%
\usepackage{amsmath,amssymb,amsfonts}%
\usepackage{siunitx}
\usepackage{multirow}%
\usepackage{dcolumn}
\usepackage{ulem}
\usepackage{lineno}
\usepackage[bottom]{footmisc}

\usepackage[
    colorlinks=true,
    citecolor=blue,
    linkcolor=blue,
    urlcolor=blue,
    pdfencoding=auto,
    pdfusetitle
]{hyperref}

\begin{document}


\title{
Measurement of Neutrino Oscillation Parameters at JUNO and \\
Indication of the Neutrino Mass Ordering}




\author{Thomas Adam}
\affiliation{IPHC, Universit\'{e} de Strasbourg, CNRS/IN2P3, F-67037 Strasbourg, France}

\author{Timo Ahola}
\affiliation{University of Jyvaskyla, Department of Physics, Jyvaskyla, Finland}

\author{Nurbakyt Amanbek}
\affiliation{GSI Helmholtzzentrum f\"{u}r Schwerionenforschung GmbH, Planckstr. 1, D-64291 Darmstadt, Germany}
\affiliation{Institute of Physics and EC PRISMA$^+$, Johannes Gutenberg Universit\"{a}t Mainz, Mainz, Germany}

\author{Fengpeng An}
\affiliation{Sun Yat-sen University, Guangzhou, China}

\author{Jo\~{a}o Pedro Athayde Marcondes de Andr\'{e}}
\affiliation{IPHC, Universit\'{e} de Strasbourg, CNRS/IN2P3, F-67037 Strasbourg, France}

\author{Costas Andreopoulos}
\affiliation{The University of Liverpool, Department of Physics, Oliver Lodge Laboratory, Oxford Str., Liverpool L69 7ZE, UK, United Kingdom}

\author{Giuseppe Andronico}
\affiliation{INFN Catania and Dipartimento di Fisica e Astronomia dell Universit\`{a} di Catania, Catania, Italy}

\author{Nikolay Anfimov}
\affiliation{Joint Institute for Nuclear Research, Dubna, Russia}

\author{Vito Antonelli}
\affiliation{INFN Sezione di Milano and Dipartimento di Fisica dell Universit\`{a} di Milano, Milano, Italy}

\author{Tatiana Antoshkina}
\affiliation{Joint Institute for Nuclear Research, Dubna, Russia}

% ...
\collaboration{The JUNO Collaboration}
\email[]{juno\_pub\_comm@ihep.ac.cn}



\begin{abstract}
We report an improved measurement of neutrino oscillation parameters at the Jiangmen Underground Neutrino Observatory (JUNO) and the consequent indication of neutrino mass ordering.
The analysis is based on 207.2 live days of data, comprising 8,294 inverse beta-decay candidates collected at an average reactor baseline of 52.5\,km. A world-leading precision of approximately 1\% was achieved for the atmospheric neutrino mass splitting $\Delta m^2_{31}$, with $\Delta m^2_{31}=+2.509^{+0.027}_{-0.025}\times10^{-3}\,\mathrm{eV}^2$ for normal ordering and $\Delta m^2_{31}=-2.482^{+0.025}_{-0.026}\times10^{-3}\,\mathrm{eV}^2$ for inverted ordering, using the Daya Bay constraint in the fit to resolve the correct physical minimum under each mass-ordering hypothesis.
We further improved the world's most precise measurements of the solar oscillation parameters, obtaining $\sin^2\theta_{12}=0.3036\pm0.0064$ and $\Delta m^2_{21}=(7.388\pm0.078)\times10^{-5}\,\mathrm{eV}^2$, corresponding to relative precisions of 2.1\% and 1.1\%, respectively. 
Incorporating constraints on $\Delta m^2_{31}$ from long-baseline accelerator neutrino experiments into our analysis favored normal ordering over inverted ordering, with a one-sided significance of at least $2.1\sigma$ across all possible CP-violating phases; a Bayesian analysis yielded a Bayes factor of 10.18 in favor of normal ordering.
\end{abstract}


\maketitle

The discovery of neutrino oscillations established that neutrinos have non-zero masses and that the flavor eigenstates participating in weak interactions are mixtures of mass eigenstates~\cite{Super-Kamiokande:1998kpq,SNO:2002tuh}. In the standard three-flavor framework, neutrino oscillations are described by three mixing angles ($\theta_{12}$, $\theta_{13}$, $\theta_{23}$), two independent mass-squared differences ($\Delta m^2_{21}$ and $\Delta m^2_{31}$, with $\Delta m^2_{ij}=m^2_i-m^2_j\,(i>j)$), and the CP-violating phase ($\delta_{\rm CP}$)~\cite{ParticleDataGroup:2024cfk}. The remaining unknowns are the octant of $\theta_{23}$, the value of $\delta_{\rm CP}$ and the neutrino mass ordering (NMO), i.e., whether the mass spectrum follows the normal ordering (NO, $\Delta m^2_{31}>0$) or inverted ordering (IO, $\Delta m^2_{31}<0$).

Long-baseline accelerator~\cite{T2K:2025dwp} and atmospheric neutrino experiments~\cite{Super-Kamiokande:2023ahc,IceCube:2019dyb,KM3NeT:2021ozk} probe the NMO through matter-induced modifications of neutrino oscillations, with sensitivity that is correlated with other oscillation parameters, particularly $\delta_{\rm CP}$ and $\theta_{23}$~\cite{Esteban:2024eli}. In contrast, the Jiangmen Underground Neutrino Observatory (JUNO) probes the NMO through reactor antineutrino ($\overline{\nu}_{e}$) disappearance, exploiting the energy-dependent imprint of vacuum-dominated oscillations~\cite{Li:2025hye}, independently of $\delta{\rm CP}$ and $\theta_{23}$~\cite{Petcov:2001sy,Learned:2006wy,Zhan:2008id,Zhan:2009rs,Li:2013zyd}.

Reactor antineutrinos from the eight reactor cores at the Yangjiang and Taishan nuclear power plants, with a nominal total thermal power of 26.6~GW$_{\rm th}$, are detected in JUNO via inverse beta decay (IBD), $\overline{\nu}_{e}+p\rightarrow e^++n$~\cite{Vogel:1999zy,Strumia:2003zx}. The prompt positron signal provides an estimate of the $\overline{\nu}_{e}$ energy, while its coincidence with the delayed neutron-capture signal enables efficient background rejection. At the JUNO baseline of $\sim$52.5\,km, the prompt-energy spectrum exhibits slow oscillations governed by $\Delta m^2_{21}$, modulated by rapid oscillations arising from the interference between the $\Delta m^2_{31}$- and $\Delta m^2_{32}$-driven oscillation modes~\cite{JUNO:2015zny}. Resolving the resulting oscillation pattern requires unprecedented energy resolution, a precisely characterized detector response, a precise prediction of the unoscillated reactor-$\overline{\nu}_{e}$ spectrum, and several years of data.

The analysis presented in this paper is based on 207.2 live days of JUNO data collected between 30 August 2025 and 18 May 2026, substantially extending the 59.1-day dataset used in the initial oscillation analysis~\cite{JUNO:2025gmd}. In addition to the reactor $\overline{\nu}_{e}$ spectrum from the Daya Bay near detectors~\cite{DayaBay:2025ngb,dayabay_open_data}, this analysis incorporated the reactor $\overline{\nu}_{e}$ spectrum measured by the Taishan Antineutrino Observatory (TAO)~\cite{TAO:REF}, a high-energy-resolution satellite detector located close to a Taishan reactor core. The first TAO measurement provided a near-detector reference of the unoscillated reactor spectrum, based on 31.6 reactor-on days and 29.6 reactor-off days with IBD candidate rates of approximately 2900 and 419 events per day, respectively. Three independent groups performed parallel analyses with consistent results; the results presented here are from one of these groups. A blinding strategy was followed in which the analysis framework---including event selection, detector response, background estimation, and systematic uncertainty treatment---was validated before performing the fit and examining the results. The reactor operational inputs were unblinded only at the final stage.

In this paper we report improved measurements of $\sin^2\theta_{12}$ and $\Delta m^2_{21}$, together with a first measurement of $\Delta m^2_{31}$ using JUNO data. Due to the limited size of the current dataset, the JUNO-only $\chi^2$ profile exhibits multiple degenerate local minima in $\Delta m^2_{31}$. Incorporating the Daya Bay measurement~\cite{DayaBay:2022orm} of $|\Delta m^2_{31}|$ as a constraint in the fit, a compatible minimum was identified, with the ultimate precision driven by the JUNO data. Combined with the NuFIT long-baseline $\Delta m^2_{31}$ constraints~\cite{Esteban:2024eli}, JUNO provides an indication in favor of NO.

JUNO is located beneath the Dashi Hill in Guangdong, China, with a 650\,m rock overburden (about 1800\,m water equivalent). The central detector (CD) consists of 20\,kt of liquid scintillator (LS) contained in a spherical acrylic vessel (AV) and instrumented with about 17,596 20-inch and 25,587 3-inch photomultiplier tubes (PMTs). It is surrounded by a Water Pool (WP) Cherenkov detector that provides shielding and an active cosmic-muon veto, while an external plastic-scintillator Top Tracker further improves muon-track reconstruction. The measured attenuation lengths of LS and ultra-pure water were ($22.2\pm1.0$)\,m and ($80.6\pm6.4$)\,m, respectively, which are important for maintaining a high light yield. A detailed description of the JUNO detector and its performance can be found in Refs.~\cite{JUNO:2015zny,JUNO:2022hxd,JUNO:2025fpc}.

A comprehensive calibration program was performed during the data taking period. The PMT calibration constants were determined using weekly deployments of UV laser and $^{241}$Am–$^{13}$C (Am-C) sources at the detector center. Several calibration campaigns were carried out to characterize the detector non-uniformity. The automatic calibration unit deployed sources along the central axis, including $^{137}$Cs, $^{54}$Mn, $^{40}$K, and $^{60}$Co and $^{68}$Ge $\gamma$ sources, the Am-C neutron source, and a UV laser. The cable loop system enabled a two-dimensional calibration along a vertical half-plane with the Am-C, $^{68}$Ge, and $^{137}$Cs sources. This calibration extended to larger radii than in Ref.~\cite{JUNO:2025gmd}, including locations near the acrylic vessel. No $^{222}$Rn leakage or other contamination was introduced during calibration. Additionally, a three-dimensional calibration was performed using $\alpha$ particles from naturally occurring $^{214}$Po decays.

Event reconstruction relied on the photon arrival times and integrated charge extracted from PMT waveforms. A combined time- and charge-likelihood method was employed to simultaneously reconstruct the event vertex and energy, incorporating corrections for spatial non-uniformity derived from the calibration data and natural sources. For $^{68}$Ge deployed along the $z$-axis with $|z|<17.2$\,m, the vertex reconstruction bias was within $\pm$5\,cm, and the resolution ranged from 9 to 13\,cm~\cite{JUNO:2026:vertexReco}. Muon tracks were reconstructed using Transformer-based neural networks~\cite{JUNO:2026:muonReco} that combined the spatial pattern and timing of PMT hits to infer trajectories. Events were first classified as single-muon or muon-bundle events and passed to dedicated networks, which reconstructed one straight track or approximated the bundle by two parallel tracks, respectively. These tracks were then used to identify and veto cosmogenic background candidates based on their spatial and temporal correlations with preceding muons, and to evaluate the residual cosmogenic background.

\begin{figure}[t]
\centering
\includegraphics[width=0.45\textwidth]{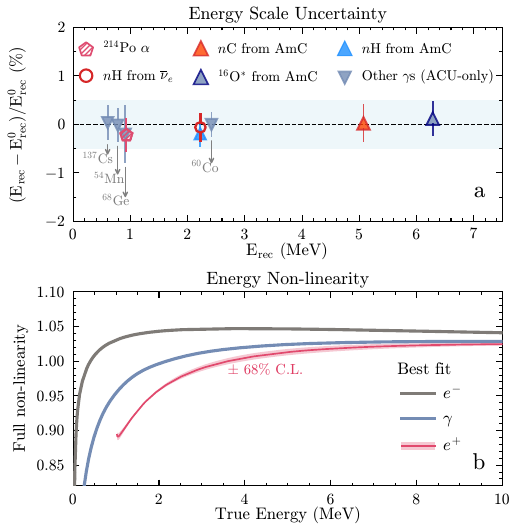}
\caption{ \textbf{(a)} Relative deviation of the reconstructed energy after all corrections from the respective detector-center reference for artificial and natural sources within the FV. Error bars cover residual spatial non-uniformity and temporal variation. ``Other $\gamma$s'' are deployed only along the vertical axis with limited coverage; for all other sources, the mean deviation and uncertainties are within the 0.5\% blue band. \textbf{(b)} Full energy scale non-linearity versus true energy for electrons, gammas, and positrons, derived from simultaneous fit to calibration sources and cosmogenic isotopes. The shaded band shows the 68\% C.L. fit uncertainty of the positron curve.}
\label{fig:energyScale}
\end{figure}

In the first calibration campaign, the light yield at the detector center was approximately 1600 p.e./MeV for the two 511\,keV annihilation $\gamma$'s from the $^{68}$Ge and 1785 p.e./MeV for the 2.223\,MeV neutron-capture $\gamma$, in good agreement with Monte Carlo (MC) simulations. Time-dependent variations (up to $\sim$3\%) were corrected using $\alpha$'s from $^{214}$Po in the full LS volume, and checked using the neutron-capture peak from IBD candidates. A residual $(r, \theta)$-dependent correction ($\sim$1\%) was applied using high-statistics, time-sliced $^{214}$Po decays to further suppress the spatial non-uniformity. After all these corrections, the energy scale was anchored at the neutron-capture peak of the total IBD candidates within the $R<$\,17.2\,m fiducial volume (FV). The final residual spatial non-uniformity and temporal variations were within 0.2\% and 0.3\%, respectively. Figure~\ref{fig:energyScale}a shows the deviation of the final reconstructed energy from the detector-center reference for all artificial and natural sources. The residual energy scale uncertainty, accounting for all sources within the FV, was less than 0.5\% (Fig.~\ref{fig:energyScale}a). 

%
The energy non-linearity model accounts for three distinct effects: ionization quenching, Cherenkov light emission, and instrumental non-linearity. The physics inputs for quenching and Cherenkov emission were informed by Geant4 simulations. Small PMTs were used to constrain the instrumental non-linearity of the 20-inch PMTs stemming from intrinsic PMT effects, front-end electronics, and imperfect charge reconstruction. The overall model parameters were determined by a simultaneous data-driven fit to dedicated calibration data from $\gamma$ sources at the detector center from 0.5 to 6\,MeV ($^{137}$Cs, $^{54}$Mn, $^{40}$K, $^{60}$Co, $^{68}$Ge, and Am-C), supplemented by the continuous $\beta$ spectra of the cosmogenic isotopes $^{12}$B ($\beta^-$, up to $\sim$14\,MeV) and $^{11}$C and $^{10}$C ($\beta^+$, up to $\sim$2\,MeV) distributed throughout the FV. The resulting non-linearity curves for electrons, gammas and positrons are shown in Fig.~\ref{fig:energyScale}b. The positron curve used directly in the oscillation analysis was constrained to 0.6\% precision across the relevant energy range. 

\begin{figure}[b]
\centering
\includegraphics[width=0.45\textwidth]{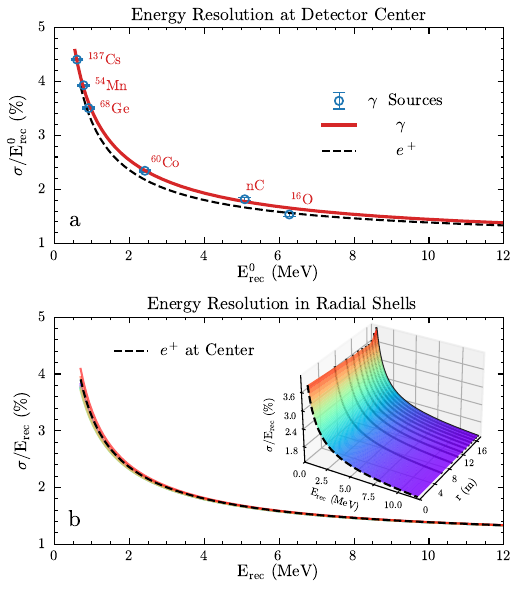}
\caption{\textbf{(a)} Energy resolution vs. reconstructed energy at the detector center. Data points are $\gamma$-source measurements; the red curve is the $\gamma$-resolution fit, and the dashed black curve is the detector-center $e^+$ resolution model. \textbf{(b)} Radial dependence of the energy resolution as a function of reconstructed energy. Colored curves represent the $e^+$ resolution model for individual radial shells, from the innermost (blue) to the outermost (red). The inset illustrates the full 3D dependence of $\sigma/E$ on reconstructed energy and radius.
}
\label{fig:energyRes}
\end{figure}

%
The energy resolution for the $^{68}$Ge $\gamma$-rays at the detector center was approximately 3.5\% ($\sigma/E$), slightly worse than the MC prediction~\cite{JUNO:2024fdc}.
The $\gamma$-resolution curve shown in Fig.~\ref{fig:energyRes}a was obtained by fitting calibration data from $\gamma$ sources placed at the detector center. From these same data, the detector-center $e^+$ resolution model was subsequently derived, using a Geant4-based correction that accounts for the $\gamma$-to-$e^{+}$ resolution difference. The radial dependence of the energy resolution, which varied by less than 5\% relative to the center, was obtained using a high-statistics $^{214}$Po sample distributed throughout the FV. The resolution improved slightly at intermediate radii owing to enhanced photon collection, then degraded near the acrylic boundary due to total internal reflection. This radial variation, assumed to be energy-independent, was combined with the $e^{+}$ resolution at the detector center to model the position-dependent energy resolution used to predict the reactor $\overline{\nu}_{e}$ energy spectrum (Fig.~\ref{fig:energyRes}b), extending the treatment used in the previous analysis.

%
An energy response matrix was constructed for the oscillation analysis to map deposited energy to reconstructed energy, incorporating the detector response components described above. Specifically, the matrix incorporated the following effects: (1) the positron energy non-linearity, constrained by calibration sources and cosmogenic isotopes; (2) the radius-dependent energy resolution and energy scale, modeled via volume-weighted averaging of response matrices for individual radial shells; (3) a $^{14}$C pile-up correction applied via spectral convolution with the simulated $^{14}$C $\beta^-$ spectrum. The pile-up probability ($\sim$5\%) was determined \textit{in situ} by fitting the $^{137}$Cs calibration spectrum and was cross-checked via a machine-learning analysis of periodically triggered PMT hit-multiplicity data~\cite{JUNO:2026:C14Activity}. This was consistent with an independent screening campaign using a low-background setup at the JUNO site, where two PMTs measured the energy spectrum of a one-liter LS sample to determine its $^{14}$C content. To propagate the uncertainties of these effects, a spectral covariance matrix was built via Monte Carlo sampling and was added to the JUNO systematic covariance matrix, introduced later.

%
IBD candidates were selected according to the criteria summarized in Table~\ref{tab:selection}, with the improvements relative to Ref.~\cite{JUNO:2025gmd} indicated by daggers. PMT flashers, an instrumental background caused by spontaneous internal light emission, were efficiently rejected using the standard deviations of the PMT hit multiplicity and timing distributions. Extending the FV radius from $R<16.5$\,m to $R<17.2$\,m improved the detection efficiency by $\sim$9\%. Including the 4.95\,MeV $\gamma$ ray from neutron capture on $^{12}$C in the delayed-energy selection raised the signal yield by approximately 0.7\%. The multiplicity cut was refined accordingly by extending the energy range of additional events, thereby ensuring an unambiguous identification of isolated IBD candidates. The muon veto was enhanced with muon track information to further suppress long-lived cosmogenic $^9$Li/$^8$He. A likelihood-ratio cut was implemented to reject accidental background. The dominant backgrounds remain cosmogenic $^9$Li/$^8$He, geoneutrinos, antineutrinos from distant reactors, and $^{214}$Bi-$^{214}$Po cascades; minor backgrounds were either intrinsically small or strongly suppressed by selection criteria. Detailed background studies are presented below and summarized in Table~\ref{tab:ibd}.

\begin{table}[b]
\caption{Summary of the IBD event selection criteria. The subscript $p$ ($d$) denotes the prompt (delayed) signal; $E$ the energy; $spn$ the spallation neutrons; $\mu$ the cosmic muons; $Q$ the total charge; $\Delta T$ the time interval; $\Delta R$ the distance, and in the subscript ``-'' denotes ``to'', e.g. $d-p$ stands for delayed-to-prompt.}
\label{tab:selection}
\footnotesize
\begin{ruledtabular}
\begin{tabular}{ll}
\multicolumn{2}{c}{\bf Muon ($\mu$) and spallation neutron ($spn$) identification}\\ \hline
$\mu$  & $Q>3 \times 10^4$\,pe (CD), or $Q>700$\,pe (WP) \\[0.1cm]
$spn$ & $Q\in (3, 30)\times10^3$\,pe, $\Delta T_{spn-\mu}\in$(0.02, 2)\,ms \\[0.2cm]

\multicolumn{2}{c}{\bf IBD selection criteria} \\
\hline
PMT flasher removal & Using hit multiplicity and timing info \\ [0.1cm]

Prompt $E$ &
$0.7<E_p<12.0$ MeV \\ [0.1cm]

Delayed $E$ &
$2.0<E_d<2.5$ MeV ($n$ capture on H) \\
& $4.5<E_d<5.5$ MeV ($n$ capture on $^{12}$C) $^\dagger$ \\[0.1cm]

Coincidence &
$5<\Delta T_{{p-d}}<1000~\mu$s,\quad $\Delta R_{p-d}<1.5$\,m \\[0.1cm]

Fiducial volume &
$r_p<17.2$\,m $^\dagger$, \quad $|z_p|<15.5$\,m \\[0.1cm]

Muon veto &
$\Delta T_{d-\mu}>5$\,ms (CD), $\Delta T_{d-\mu}>2$\,ms (WP)\\
& $\Delta R_{d-spn}>4$ \,m or $\Delta T_{d-spn}>1.2$\,s \\
& $\displaystyle
\dfrac{\Delta R_{d-\mu}}{2.5~{\rm m}}+
\dfrac{\Delta T_{d-\mu}}{0.5~{\rm s}}>1$ $^\dagger$\\[0.3cm]

Multiplicity & No additional event with \\
& $E>0.7$ MeV (inside the FV) nor \\
& $2.0<E<6.0$ MeV (outside the FV) $^\dagger$ \\
& occurred within $(t_d-2,\,t_d+1)$\,ms \\[0.1cm]

L.L.R. & Likelihood-ratio selection $^\dagger$\\
\end{tabular}
\end{ruledtabular}
\parbox{\columnwidth}{\raggedright
\scriptsize
$^\dagger$\, Improved selections with respect to Ref.~\cite{JUNO:2025gmd}.}
\end{table}

%
The dominant background from long-lived cosmogenic $^9$Li/$^8$He $\beta$-n decays is suppressed via spallation neutron and $\mu$-track veto. The total (pre-veto) and residual (post-veto) background rates were obtained by fitting the time and lateral-distance distributions of IBD candidates relative to the last preceding cosmic muon. The fits were performed simultaneously in subsamples binned by tagged muon visible energy. In each subsample, the reference muon was selected independently.
The He-to-Li production ratio was constrained to $0.0143\pm0.0023$ using Daya Bay measurements~\cite{DayaBay:2024xye}. Off-window events matched to each tagged muon were used to construct a template for muon-uncorrelated IBD candidates; subtracting this template from the in-window event sample, the excess at small distances yielded an independent estimate of residual $^9$Li/$^8$He, which was included as an additional constraint in the time-distribution fit. The larger data sample compared with~\cite{JUNO:2025gmd} enabled more precise normalization and spectral shape determination. A 26\% normalization uncertainty was assigned to account for fitting errors and data--simulation discrepancies in the dependence on visible muon energy. The spectral-shape uncertainty was constrained to 20\% by comparisons among spallation-neutron-tagged $^9$Li/$^8$He events, nuclear database calculations, and a polynomial-parameterized model validated by Ref.~\cite{DayaBay:2022orm}.

Expanding the FV increased accidental backgrounds. To suppress these, a likelihood ratio was constructed for each IBD candidate using probability density functions (PDFs) derived from signal-like events ($E_p>3$\,MeV; above natural radioactivity) and an off-window accidental sample. These PDFs incorporated prompt and delayed spatio-temporal distributions, distance to the AV surface, and delayed energy. This cut reduced the accidental backgrounds by a factor of $\sim$40, with only a $(1.00 \pm 0.15)$\% signal inefficiency.

The $\beta$-$\alpha$ cascades from $^{214}$Bi/$^{214}$Po (radon daughters) introduce correlated backgrounds, in which delayed $\alpha$ signals can leak into the neutron-capture energy region due to $\gamma$'s from excited $^{214}$Po states~\cite{Borexino:2019gps} or proton elastic scattering~\cite{SNO:2025koj}. Exploiting the higher $^{222}$Rn concentration in the bottom hemisphere, we extracted the delayed-energy spectrum of $^{214}$Bi/$^{214}$Po using bottom-minus-top subtraction. This model-independent approach estimated that $(0.021 \pm 0.015)\%$ of $^{214}$Po signals leak into the neutron-capture energy region, causing the associated $^{214}$Bi/$^{214}$Po decays to mimic IBD events.

\begin{table}[b]
    \caption{
    Summary of signal and background rates in counts per day (cpd). Pre-fit rates with constraint uncertainties are inputs to the oscillation fit; post-fit rates are the best-fit values. All rates are as observed, not efficiency-corrected.}
    \label{tab:ibd}
    \centering
    \footnotesize
    \setlength{\tabcolsep}{4pt}
    \renewcommand{\arraystretch}{1.2}
    \begin{ruledtabular}
    \begin{tabular}{lcc}
    DAQ live time (days) & \multicolumn{2}{c}{207.2} \\
    IBD candidates & \multicolumn{2}{c}{8294} \\ 
    \hline 
    \textbf{Signal rates (cpd)} & Pre-fit & Post-fit \\
    \quad Reactor $\overline{\nu}_{e}$ & 36.02 $\pm$ 0.81  &  34.62 $\pm$ 0.64  \\ 
    {\bf Backgrounds (cpd)}  & &  \\ 
    \quad ~$^{9}$Li/$^{8}$He & 1.05 $\pm$ 0.27 & 1.41 $\pm$ 0.26 \\
    \quad Geoneutrinos      &  1.30 $\pm$ 0.52 $^\dagger$ & 2.12 $\pm $ 0.32 \\
    \quad World reactors    & 0.92 $\pm$ 0.09 & 0.92 $\pm$ 0.09 \\
    \quad $^{214}$Bi-$^{214}$Po & 0.42 $\pm$ 0.31 & 0.57 $\pm$ 0.34 \\
    \quad $^{13}$C($\alpha$, n)$^{16}$O & 0.05 $\pm$ 0.02 & 0.05 $\pm$ 0.02 \\
    \quad Fast neutrons         & 0.02 $\pm$ 0.02 & 0.02 $\pm$ 0.02 \\
    \quad Double neutrons       & 0.10 $\pm$ 0.10 & 0.13 $\pm$ 0.10 \\
    \quad Atmospheric neutrinos & 0.09 $\pm$ 0.04 & 0.11 $\pm$ 0.05 \\
    \quad Accidentals           & 0.07 $\pm$ 0.01 & 0.07 $\pm$ 0.01 \\ 
    \quad Total backgrounds     & 4.01 $\pm$ 0.68  & 5.40 $\pm$ 0.47  \\
    \end{tabular}
    \end{ruledtabular}
    \parbox{\columnwidth}{\raggedright
    \scriptsize 
    $^\dagger$\ Average model prediction from~\cite{JUNO:2025sfc}; unconstrained in the fit.}
\end{table}

Geoneutrinos produced in the $\beta$-decays of the $^{238}$U and $^{232}$Th decay chains inside the Earth are indistinguishable from reactor antineutrinos on an event-by-event basis. Geological models predict 1.1--2.4 IBD events per day in the LS at 100\% detection efficiency~\cite{JUNO:2025sfc}. In this analysis, the geoneutrino rate was left unconstrained in the fit, while the $^{232}$Th-to-$^{238}$U ratio was constrained to the value 0.29 with 10\% uncertainty, covering variations among geological models~\cite{JUNO:2025sfc}. A 5\% shape uncertainty was assigned to account for spectral model differences~\cite{Enomoto,Li:2024mqz}, oscillation-pattern distortions from the different spatial distributions of U and Th, and oscillation-parameter uncertainties.
Using 207.2 live days of data, the best-fit rate was $(2.12\pm0.32)$\,cpd, corresponding to an efficiency-corrected total flux of ($73\pm11$)\,TNU\footnote{One terrestrial neutrino unit (TNU) corresponds to one IBD event over 1 year exposure of $10^{32}$ free target protons at 100\% detection efficiency, see Ref.~\cite{Mantovani:2003yd}.}. 

The world-reactor background originated from distant reactors located more than 450\,km away from JUNO. Using quarterly reactor power data from Chinese nuclear plants~\cite{CNEA} and the global PRIS database~\cite{PRIS}, the expected IBD rate at JUNO was 1.24\,events/day at 100\% detection efficiency. This was slightly lower than in Ref.~\cite{JUNO:2025gmd} due to updated operational data. Rate and shape uncertainties of 10\% and 5\%, respectively, were assigned to conservatively account for uncertainties in the operational evolution of these reactors during periods for which complete information was unavailable. The minor backgrounds, namely double neutrons, fast neutrons, $^{13}$C($\alpha$, n)$^{16}$O, and atmospheric neutrinos were evaluated with the same approach as in Ref.~\cite{JUNO:2025gmd}.

Table~\ref{tab:effUncer} summarizes the absolute efficiencies and relative uncertainties for the predicted antineutrino signal. These quantities serve as inputs to the oscillation fit.

The efficiency for the enlarged FV selection was estimated as the ratio of selected high-energy ($>$3.5\,MeV) IBD candidates within the FV to the expected total in the full LS volume. The latter was estimated from the measured IBD density in an inner control region. The total uncertainty combined the statistical uncertainty with a systematic component evaluated by varying the inner normalization region. The target-proton number and its uncertainty were taken from Ref.~\cite{JUNO:2025gmd}.

For the extended FV, the prompt-delayed coincidence efficiency accounts for energy leakage effects assessed via AmC calibration and MC simulations, with additional uncertainties reflecting vertex reconstruction bias near the boundary. The efficiencies for the $\mu$ veto and multiplicity cut were determined from the corresponding loss of exposure with negligible uncertainty. The detector volume was segmented into 40-cm cubic voxels, and the accumulated livetime in each was summed to properly account for overlapping veto regions and position-dependent veto durations, leading to negligible uncertainty. The likelihood-ratio cut efficiency was derived from a data-driven approach using selected high-energy IBD candidates, with uncertainties dominated by statistical errors and PDF parameterization systematics. 

The reference reactor flux was based on the Huber--Mueller prediction~\cite{Huber:2011wv,Mueller:2011nm}, with corrections for the spectral distortion around 5\,MeV using Daya Bay data~\cite{DayaBay:2025ngb}, as well as for fission fractions, spent nuclear fuel (SNF), and non-equilibrium long-lived isotopes on a per-core basis. For the oscillation fit, the spectrum was modulated by free normalization parameters in consecutive energy segments. These parameters were constrained by a joint fit to TAO~\cite{TAO:REF} and Daya Bay near-detector data~\cite{DayaBay:2025ngb}, with an independent response model for each detector; The predicted spectrum for each detector was obtained by folding the unoscillated reactor flux with the three-flavor survival probability, the IBD cross section, the energy response matrix, and the detection efficiency. The reactor-related uncertainties are summarized in the lower part of Table~\ref{tab:effUncer}, as detailed in Ref.~\cite{DayaBay:2021dqj,DayaBay:2025ngb}. 

\begin{table}[!b]
    \caption{Summary of absolute efficiencies and their uncertainties, along with reactor-related uncertainties.}
    \label{tab:effUncer}
    \centering
    \footnotesize
    \setlength{\tabcolsep}{4pt}
    \renewcommand{\arraystretch}{1.2}
    
    \begin{ruledtabular}
    \begin{tabular}{lcc}
        \multicolumn{3}{c}{Detector and IBD selection} \\
        \hline
         & Efficiency & $\sigma_{\text{rel}}$ \\
        Target protons ($1.442\times10^{33}$)           &  --  & 1.0\% \\
        Fiducial volume          & 87.12\%      & 0.96\% \\ 
        PMT flasher rejection    & $>$ 99.9\%   & Negligible \\
        $\mu$ veto               & 93.67\%      & 0.02\% \\
        Prompt-delayed coinc.    & 95.50\%      & 0.32\% \\
        Multiplicity cut         & 95.94\%      & Negligible \\
        Likelihood-ratio cut & 98.99\%      & 0.15\% \\ 
        \hline
        Combined  & 74.01\% $^*$ & 1.4\% \\
      \hline \hline
      \multicolumn{3}{c}{Reactor flux and spectrum prediction} \\
      \hline
        Reference spectrum  $^\dagger$  &  & 1.2\% \\
        Thermal power  $^\dagger$       &  & 0.5\% \\
        Fission fraction  $^\dagger$    &  & 0.6\% \\
        Spent nuclear fuel $^\ddagger$  &  & 0.3\% \\
        Non-equilibrium  $^\ddagger$    &  & 0.3\% \\
        Different fission fractions $^\ddagger$ &  & 0.1\% \\
    \end{tabular}
    \end{ruledtabular}
    \parbox{\columnwidth}{\raggedright
    \scriptsize
    $^*$\ Pre-fit value. Post-fit: 74.07\%, with the same uncertainty.\\
    $^\dagger$\ Correlated among all reactor cores.\\
    $^\ddagger$\ Correlated among cores in the same power plant and uncorrelated across different power plants.}
\end{table}

\begin{figure}[b]
\centering
\includegraphics[width=0.5\textwidth]{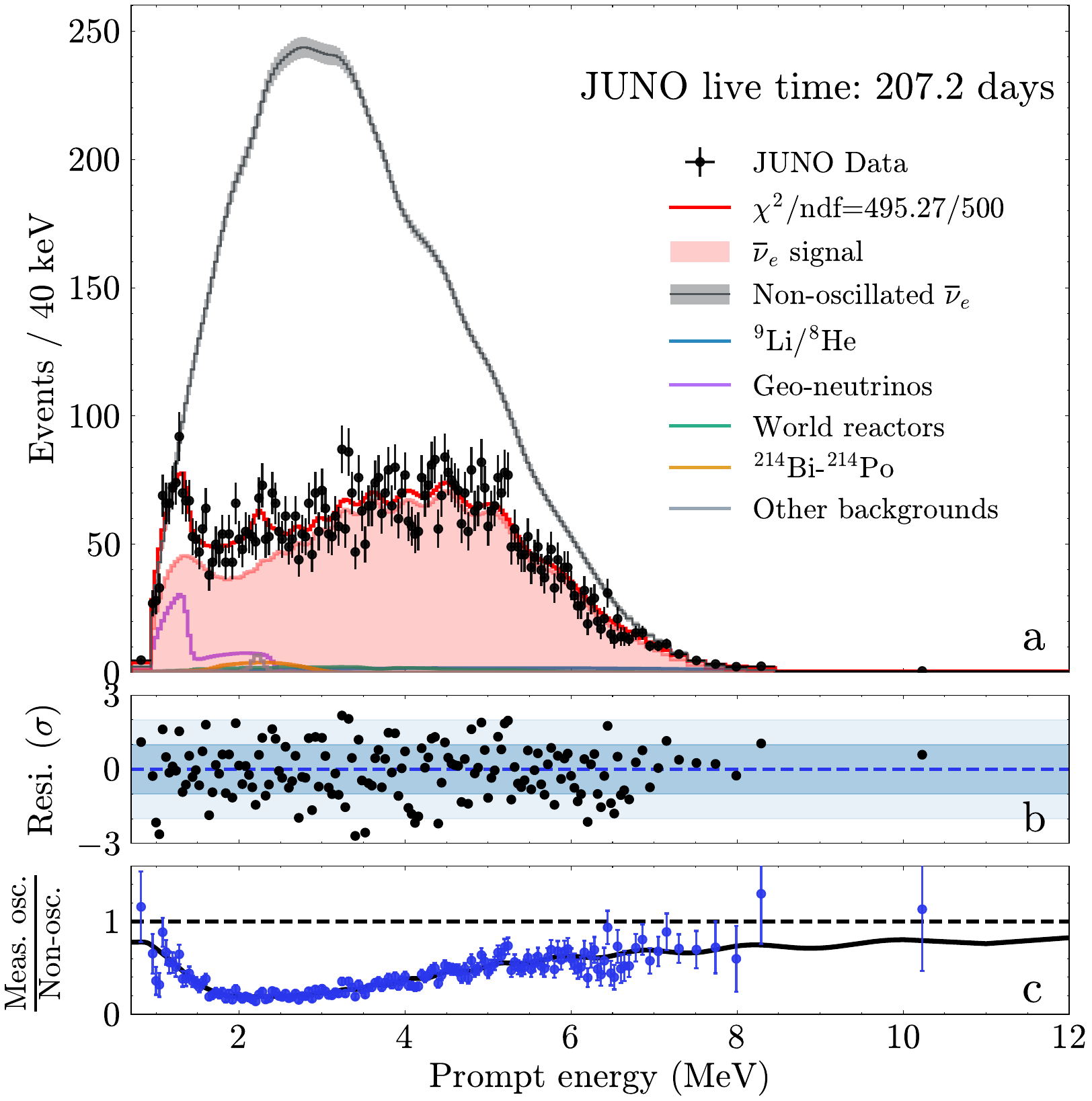}
\caption{{\bf Measured IBD prompt energy spectrum}, with all components shown at their post-fit values. \textbf{(a)} Black points: measured data with statistical uncertainties; red curve: best-fit total prediction. Shaded red: expected antineutrino signal; black solid curve and shaded band: non-oscillated reactor spectrum and its post-fit systematic uncertainty. Solid color lines: backgrounds. \textbf{(b)} Residuals between the data and the complete model. \textbf{(c)} Ratio of background-subtracted oscillated spectrum to non-oscillation prediction.}
\label{fig:ibdSpectrum}
\end{figure}

\begin{figure*}[t]
\centering
\includegraphics[width=0.75\textwidth]{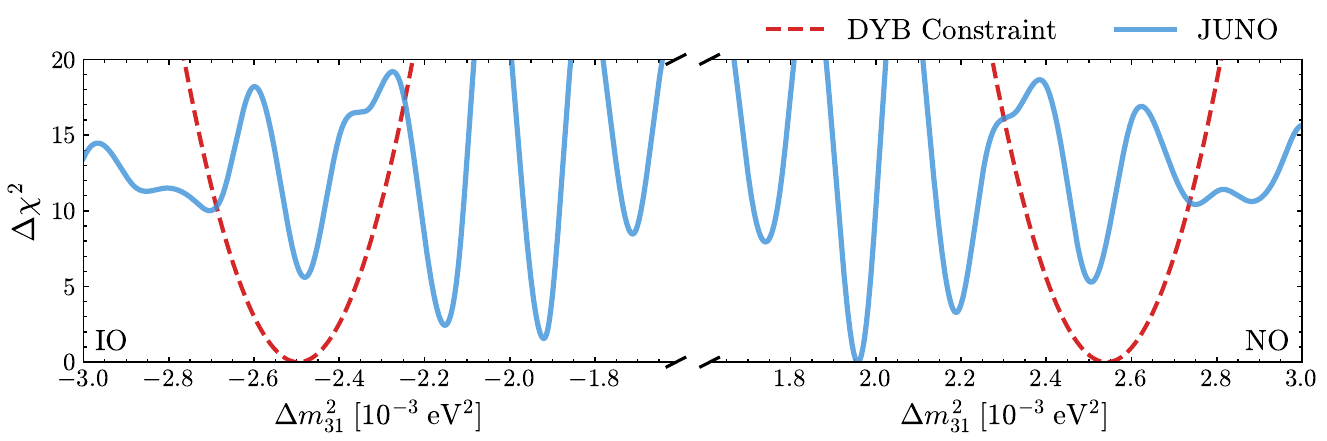}
\caption{JUNO $\Delta\chi^2$ profiles for $\Delta m^2_{31}$ (IO: left; NO: right). The Daya Bay constraints~\cite{DayaBay:2022orm} (red dashed) identify the compatible JUNO minima, while the precision within each minimum is largely determined by the JUNO profile.}
\label{fig:dm31_DYB}
\end{figure*}

%
The final JUNO dataset comprised 8,294 identified IBD candidates. 
Oscillation parameters were extracted via a joint frequentist binned-$\chi^2$ fit to the JUNO data, 90,280 IBD candidates from TAO~\cite{TAO:REF}, and Daya Bay near-detector data~\cite{DayaBay:2025ngb}. Figure~\ref{fig:ibdSpectrum} shows the reconstructed prompt-energy spectrum and the best-fit prediction, including all spectral components at their post-fit values, where the non-oscillated reactor spectrum incorporates the fit-induced modulation.

The aforementioned binned-$\chi^2$ is defined as:
\begin{multline}
\label{eqn:chi2}
\chi^{2}(\vec{p},\vec{\eta},\vec{\alpha})
= \sum_{X} (\mu^{X} - D^{X})^{T} V^{-1}_{X} (\mu^{X} - D^{X}) \\
+ \chi^{2}_{\text{pen}}(\vec{\eta})
+ \chi^{2}_{\text{pull}}(\sin^{2}\theta_{13}),\nonumber
\end{multline}
where $X$ denotes the dataset (JUNO, TAO, and Daya Bay); $\mu^X(\vec{p},\vec{\eta},\vec{\alpha})$ ($D^X$) the predicted (observed) prompt-energy spectrum in detector $X$; and $V_X$ the corresponding covariance matrix incorporating both statistical and certain systematic uncertainties for each detector. The vector $\vec{p}$ contains the oscillation parameters $\sin^2\theta_{12}$, $\Delta m^2_{21}$, $\sin^2\theta_{13}$, and $\Delta m^2_{31}$; $\vec{\eta}$ denotes the nuisance parameters associated with backgrounds, detector efficiency, reactor flux and spectrum, detector response, etc.; $\chi^2_{\rm pen}(\vec{\eta})$ the associated penalty terms. The free spectral parameters $\vec{\alpha}$ adjust the unoscillated spectral shape. The $\chi^2_{\text{pull}}$ incorporates an external constraint on $\sin^2\theta_{13}$~\cite{DayaBay:2022orm}. For both JUNO and TAO, the Combined Neyman–Pearson (CNP) prescription~\cite{Ji:2019yca} was used to define the bin-wise statistical variances in $V_X$ in a way that reduces known biases of traditional $\chi^2$ definitions for Poisson-distributed data. The additional covariance matrix contributions were experiment-specific: $V_{\rm JUNO}$ incorporated the aforementioned energy spectral covariance matrix; $V_{\rm TAO}$ included the covariance matrix arising from non-negligible energy leakage due to small detector size; and $V_{\rm DYB}$ was the published covariance matrix~\cite{DayaBay:2025ngb} incorporating both statistical and systematic uncertainties of the Daya Bay near-detector data.

\begin{figure}[b]
\centering
\includegraphics[width=0.5\textwidth]{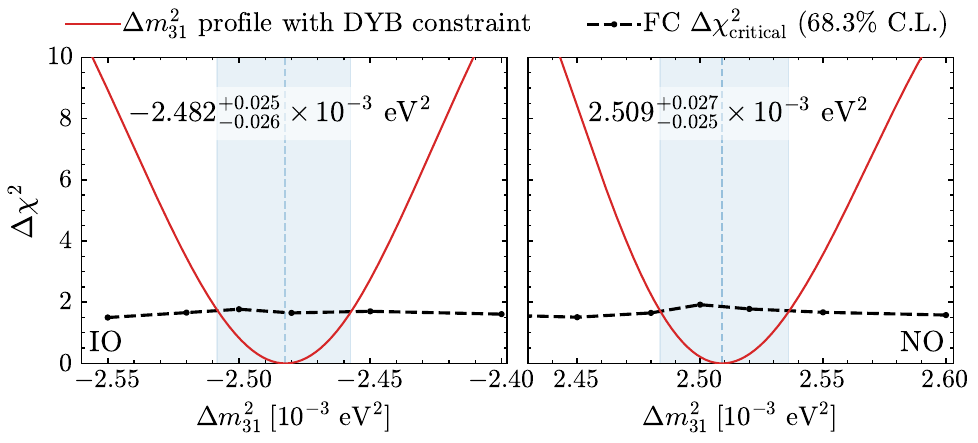}
\caption{JUNO $\Delta\chi^2$ profiles of $\Delta m^{2}_{31}$ for IO (left) and NO (right), incorporating the Daya Bay constraint~\cite{DayaBay:2022orm}. The blue vertical dashed line and band indicate the best-fit value and Feldman--Cousins (FC) 68.3\% C.L. interval, respectively. The numerical results and their $\pm1\sigma$ intervals are in the legend.}
\label{fig:dmsq31_DYB_FC}
\end{figure}

The joint reactor fit exhibited multiple local minima in the $\Delta\chi^2$ profile due to statistical fluctuations and parameter degeneracy (Fig.~\ref{fig:dm31_DYB}). To resolve this degeneracy and identify the physical minimum under each mass ordering hypothesis, the $\Delta m^{2}_{31}$ constraint from Daya Bay~\cite{DayaBay:2022orm} was incorporated as an additional pull term. To ensure proper coverage, the Feldman-Cousins method~\cite{Feldman:1997qc} was used to determine the $1\sigma$ confidence interval for $\Delta m^{2}_{31}$. The resulting measurement achieved approximately 1\% precision (Fig.~\ref{fig:dmsq31_DYB_FC}):
\begin{align}
\Delta m^2_{31} &= -2.482^{+0.025}_{-0.026} \times 10^{-3}\; {\rm eV}^2\, (\text{IO})\,,\nonumber\\
\Delta m^2_{31} &= +2.509^{+0.027}_{-0.025} \times 10^{-3}\; {\rm eV}^2\, (\text{NO})\,.\nonumber
\end{align}

\begin{figure}[b]
\centering
\includegraphics[width=0.48\textwidth]{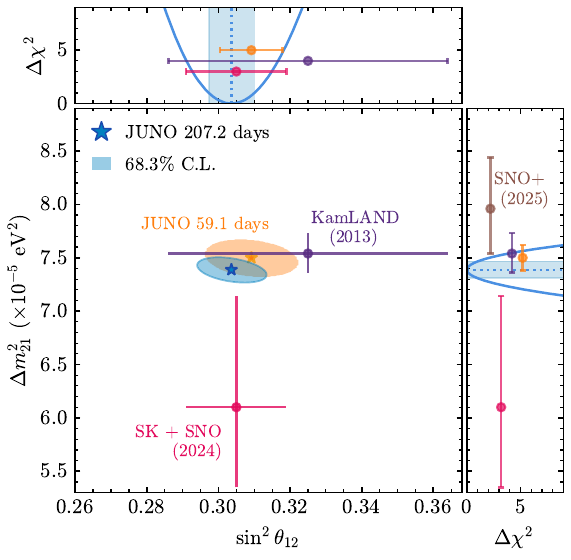}
\caption{Allowed regions in the $\sin^2\theta_{12}$--$\Delta m^{2}_{21}$ plane. Stars: JUNO best-fit values; shaded ellipses: $1\sigma$ (68.3\% C.L.) allowed regions. Previous results~\cite{JUNO:2025gmd} in orange, and new results in blue. Top and right panels show the $\Delta\chi^2$ profiles with $1\sigma$ bands (blue). For comparison, results from KamLAND~\cite{KamLAND:2013rgu}, SNO+~\cite{SNO:2025koj}, and combined SK+SNO~\cite{Super-Kamiokande:2023jbt} are included.}
\label{fig:contour}
\end{figure}

The solar parameters $\sin^2\theta_{12}$ and $\Delta m^{2}_{21}$ are insensitive to the external $\Delta m^{2}_{31}$ constraint and the mass ordering hypothesis, with the best-fit shifts below $0.05\sigma$. The allowed regions obtained with the Daya Bay $\Delta m^{2}_{31}$ constraint under NO are shown in Fig.~\ref{fig:contour}. The corresponding best-fit values are:
\begin{align}
\sin^2 \theta_{12} &= 0.3036\,\pm\,0.0064\,, \nonumber\\
\Delta m^2_{21} &= (7.388\,\pm\,0.078) \times 10^{-5}\; {\rm eV}^2\,. \nonumber
\end{align}

\begin{figure*}[t]
\centering
\includegraphics[width=0.75\textwidth]{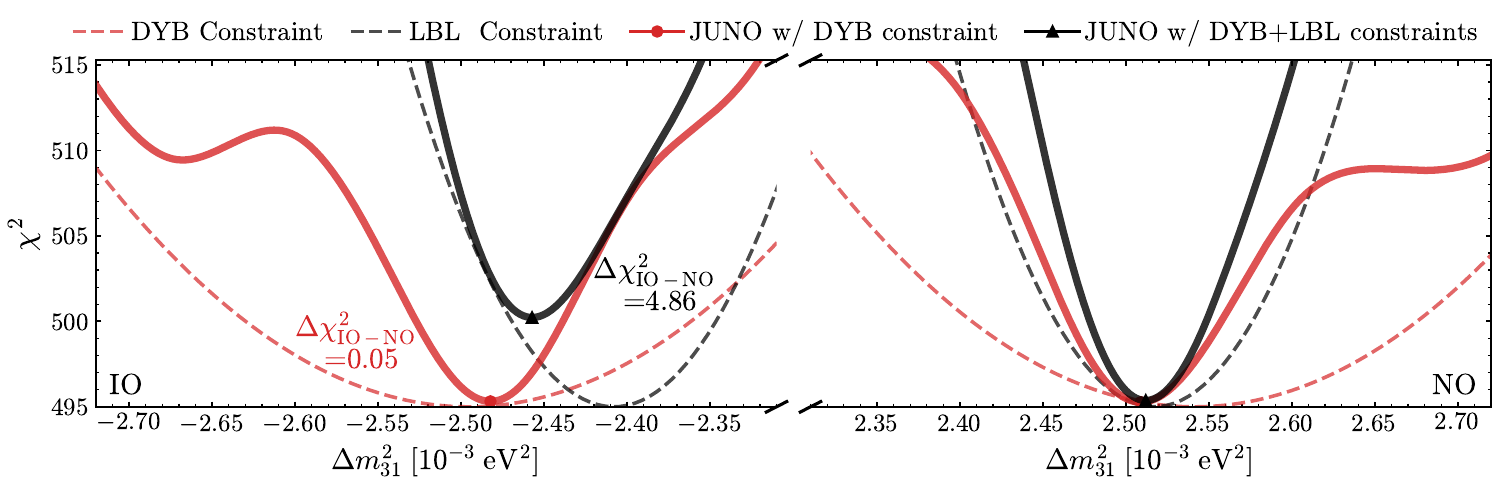}
\caption{$\chi^2$ profiles versus $\Delta m^2_{31}$ for IO (left) and NO (right). Red and black solid curves show the JUNO fit with Daya Bay constraint~\cite{DayaBay:2022orm} alone and with the additional NuFit LBL constraint~\cite{Esteban:2024eli}, respectively; dashed curves show the corresponding pull terms. Circles (DYB) and triangles (DYB+LBL) mark the best-fit values. The LBL constraint enhances the preference for NO, as quantified by $\Delta\chi^2_{\rm IO-NO}$ (values given in the IO panel). The fits constrain the unoscillated spectrum using TAO~\cite{TAO:REF} and Daya Bay near detectors~\cite{DayaBay:2025ngb}.}
\label{fig:dm31_DYB_LBL}
\end{figure*}

\begin{figure}[!h]
\centering
\includegraphics[width=0.45\textwidth]{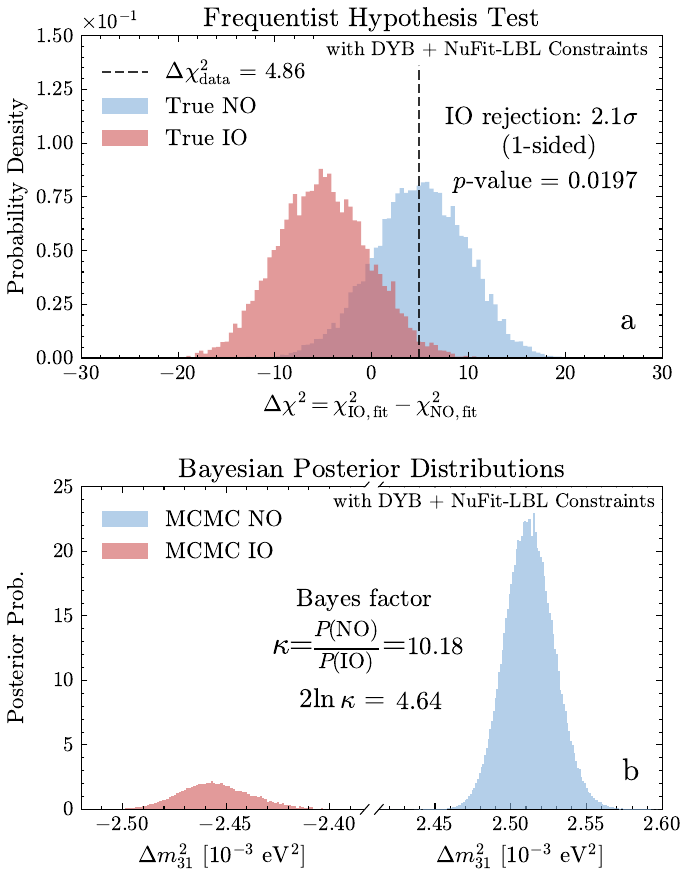}
\caption{ 
\textbf{(a)} Frequentist distributions of the test statistic $\Delta\chi^2=\chi^2_{\rm IO,fit}-\chi^2_{\rm NO,fit}$, together with the observed value (vertical dashed-dotted line).
\textbf{(b)} Bayesian posterior probability densities of $\Delta m^2_{31}$ obtained from the Markov Chain Monte Carlo analysis. Color convention: NO (blue) and IO (red).}
\label{fig:nmoSignificance}
\end{figure}

To resolve the mass ordering degeneracy, the NuFit long-baseline (LBL) result~\cite{Esteban:2024eli} was further incorporated via an additional pull term under each mass-ordering hypothesis, together with the $\Delta m^{2}_{31}$ constraint from Daya Bay, without using the absolute $\Delta\chi^{2}$ of the LBL fit. This enhanced the NMO sensitivity via the tension between the $\Delta m^{2}_{31}$ measured by reactor and LBL experiments under the NO or IO hypothesis, as discussed in Refs.~\cite{Nunokawa:2005nx,Li:2013zyd}.
The combined constraints led the fit to favor NO over IO with $\Delta \chi^2=4.86$ (see Fig.~\ref{fig:dm31_DYB_LBL}). The significance was derived using a frequentist approach~\cite{Blennow_2014}, where expected $\Delta\chi^2$ distributions were generated for each mass ordering hypothesis using a sufficiently large number of pseudo-experiments. In the simulation, the $\Delta m^{2}_{31}$ pulls from Daya Bay and LBL experiments were sampled while maintaining the proper relation between the NO and IO values. The parameter mapping depended on the oscillation parameters and, for LBL observables in particular, was further modulated by $\delta_\text{CP}$, as described by:
\begin{equation*}
\begin{aligned}
&\Delta m_{31}^{2,\,\text{IO}} = -\Delta m_{31}^{2,\,\text{NO}} + \Delta m_{21}^{2} \Big\{ 2\cos^2\theta_{12} \\
&\quad - \sin(2\theta_{12})\,\sin\theta_{13}\,\big(\tan\theta^\text{NO}_{23}\cos\delta^\text{NO} + \tan\theta^\text{IO}_{23}\cos\delta^\text{IO}\big) \Big\}.
\end{aligned}
\end{equation*}

Both this relation and its inverse were employed in the simulation. Scanning the full $\delta_\text{CP}$ range and accounting for its NO--IO correspondence, pseudo-experiments under both true hypotheses yielded the $\Delta\chi^2$ test-statistic distributions shown in Fig.~\ref{fig:nmoSignificance}a. The NMO determination is a binary hypothesis test and is therefore not subject to a look-elsewhere correction~\cite{Zhang:2026ase}. The observed value corresponded to a $p$-value of 0.0197, or a one-sided significance of $2.1\sigma$ ($2.3\sigma$ for the two-sided convention), for rejecting the IO. An independent Bayesian analysis was performed using Markov chain Monte Carlo, with a uniform prior over the two NMO hypotheses. Gaussian priors were applied to the Daya Bay $\Delta m^{2}_{31}$ and $\sin^{2}\theta_{13}$ measurements and to the LBL $\Delta m^{2}_{31}$ measurement. This analysis favored NO over IO with a Bayes factor of $\kappa=10.18$ (Fig.~\ref{fig:nmoSignificance}b). Here, $2\ln\kappa$ is on the same scale as the standard likelihood-ratio test statistic~\cite{Kass:1995loi}. The slight correlation between the Daya Bay $\Delta m^{2}_{31}$ and $\sin^{2}\theta_{13}$ measurements was found to have a negligible impact on both the frequentist and Bayesian analyses.

In summary, we presented an updated reactor antineutrino oscillation analysis based on 207.2 live days of JUNO data. For the first time, reactor antineutrino data from the TAO detector~\cite{TAO:REF} was incorporated into the oscillation analysis. The world's most precise values of $\sin^2\theta_{12}$ and $\Delta m^2_{21}$ were further improved, reaching relative precisions of 2.1\% and 1.1\%, respectively. The uncertainty breakdown showed that, although statistical uncertainty remained dominant, systematic uncertainties, particularly those associated with the Bi--Po background and detection efficiency, became significant and could be reduced in future analyses. The first JUNO measurement of $\Delta m^2_{31}$ was reported with a precision of approximately 1\%, using the Daya Bay $\Delta m^2_{31}$ constraint~\cite{DayaBay:2022orm} to select the physical minimum under each NMO hypothesis.

The precision measurement of $\theta^{}_{12}$ at JUNO provides a more stringent test of the lepton-flavor mixing pattern and may help to reveal the underlying flavor symmetries in the lepton sector~\cite{Xing:2020ijf}. Together with the precision measurements of the two neutrino mass-squared differences, it can also provide tight constraints on new-physics scenarios beyond the three-flavor neutrino-oscillation paradigm~\cite{JUNO:2015zny}.

At present, the limited JUNO statistics results in multiple local minima in $\Delta m^2_{31}$. The Daya Bay measurement~\cite{DayaBay:2022orm} identifies the relevant minimum under each NMO hypothesis, 
while JUNO determines the final precision of $\sim$1\%. Combining the JUNO measurement with the $\Delta m^2_{31}$ constraints from long-baseline neutrino experiments~\cite{Esteban:2024eli} provided the first indication of NO in a JUNO-based analysis, with a minimum one-sided significance of $2.1\sigma$ ($2.3\sigma$ for the two-sided convention) and a Bayes factor of 10.18. The data also provides the first measurement of the geoneutrino flux at the JUNO site and constituted the world's largest geoneutrino candidate sample to date. These results establish JUNO as a precision neutrino experiment and mark an important milestone toward future measurements, including a high-significance determination of the NMO.

\bibliography{junoNMO}

\end{document}